\documentclass[
    ,final            % use final for the camera ready runs
  ]
  {aipproc}

\layoutstyle{8x11single}

\usepackage{amssymb}

\begin{document}

\title{X-Raying the MOJAVE Sample of
Compact Extragalactic Radio Jets}


\author{M. Kadler}{
  address={Astrophysics Science Division, NASA's Goddard Space Flight Center, Greenbelt, MD 20771, USA}
,altaddress={NASA Postdoctoral Research Associate}
}

\author{G. Sato}{
  address={Astrophysics Science Division, NASA's Goddard Space Flight Center, Greenbelt, MD 20771, USA}
}

\author{J. Tueller}{
  address={Astrophysics Science Division, NASA's Goddard Space Flight Center, Greenbelt, MD 20771, USA}
}

\author{R. M. Sambruna}{
  address={Astrophysics Science Division, NASA's Goddard Space Flight Center, Greenbelt, MD 20771, USA}
} 

\author{C. B. Markwardt}{
  address={Astrophysics Science Division, NASA's Goddard Space Flight Center, Greenbelt, MD 20771, USA}
}

\author{P.~Giommi}{
  address={ASI Science Data Center, ESRIN, I-00044 Frascati, Italy}
}

\author{N. Gehrels}{
  address={Astrophysics Science Division, NASA's Goddard Space Flight Center, Greenbelt, MD 20771, USA}
}   
 
\begin{abstract}
The MOJAVE sample is the first large radio-selected, VLBI-monitored AGN sample 
for which complete X-ray spectral information is being gathered. We report on 
the status of \textit{Swift} survey observations which complement the available 
archival X-ray data at 0.3-10\,keV and in the UV with its XRT and UVOT 
instruments. Many of these 133 radio-brightest AGN in the northern sky are now 
being observed for the first time at these energies. These and complementary
other multi-wavelength observations provide a large statistical
sample of radio-selected AGN whose spectral energy distributions
are measured from radio to gamma-ray wavelengths, available at the 
beginning of GLAST operations in 2008.
Here, we report the
X-ray spectral characteristics of 36 of these previously unobserved
MOJAVE sources. In addition, the number of 
MOJAVE sources detected by the BAT instrument in the hard X-ray band is 
growing: we report the detection of five new blazars with BAT. 
\end{abstract}

\maketitle

%%%%%%%%%%%%%%%%%%%%%%%%%%%%%%%%%%%%%%%%%%%%
%% MAINMATTER
%%%%%%%%%%%%%%%%%%%%%%%%%%%%%%%%%%%%%%%%%%%%

\section{X-Ray Observations of MOJAVE and Relevance for GLAST}
Radio-loud core-dominated (RLCDs) active galactic nuclei (AGN) are an important class of 
extragalactic supermassive black-hole systems 
whose bright compact radio cores can be imaged at milliarcsecond resolution with Very-Long-Baseline 
Interferometry (VLBI). The vast majority of all RLCDs are blazars, BL Lac objects and flat-spectrum 
radio quasars, but also some Seyfert and broad-line radio galaxies exhibit bright compact radio cores 
with powerful relativistic jets. 
Since 1994, the VLBA 2\,cm Survey \cite{Kel04} and its 
follow-up MOJAVE (\cite{Lis05}, and Lister 2007, these proceedings) have 
been monitoring the structural changes in the jets of more than 200 AGN with bright compact radio 
cores. At present, the MOJAVE monitoring regularly observes 192 objects, including a flux-density
limited, radio-selected, statistically complete sub sample of the 133 radio-brightest compact
extragalactic jets in the northern sky: the MOJAVE sample\footnote{See the program website for
a detailed description of the selection criteria and a regularly updated list of the full sample: {\tt http://www.physics.purdue.edu/astro/MOJAVE/}}. 

Compact radio jets of blazars, typically show ejections of relativistic plasma every few months to years related to 
the formation of new jet component that travel at superluminal speeds of up to $\sim 30$\,$c$ down the jet.
On the other hand, blazars are known to be bright and rapidly variable gamma-ray sources.  
The connection of jet-formation events to the broadband-SED variability of blazars is 
an important aspect of both the MOJAVE and GLAST project and calls for multi-wavelength coordination with other 
large surveys at intermediate observing wavelengths. 
In order to be able to react \textit{swiftly} to gamma-ray flares detected by GLAST, to plan and coordinate 
multiwavelength observations, it is important to gather broadband SED information about the objects of interest
in advance to the arrival of first GLAST data. In our case of a radio-selected AGN sample that is continuously
monitored with VLBI and single-dish telescopes (see Fuhrmann et al., these proceedings), \textit{Swift} with its optical/UV (UVOT) and X-ray
(XRT) capabilities, is ideally suited to provide these SED data to connect the radio and gamma-ray bands.  

\subsection{\textit{Swift} X-Ray Survey of the MOJAVE Sample}
An archival X-ray spectral survey of all publicly 
available data for MOJAVE sources 
has been conducted 
by \cite{Kad05}. This so-called 2cm-X-Sample currently 
contains 50 out of 133 MOJAVE sources. 17 additional sources have been observed by 
\textit{Swift} before the beginning of our project. 
This still leaves us with 66 of the radio brightest blazars in the northern sky having never been observed with 
an X-ray spectroscopic mission above 2\,keV. 

We have started a program to observe all the so-far unobserved MOJAVE sources 
with the \textit{Swift} XRT as part of the \textit{Swift} blazar key project. This will produce the first statistically 
complete large radio-selected sample of blazars and other RLCDs. The observations will complete 
the data base of the 2\,cm-X-Sample of photon indices, source-intrinsic absorbing column densities 
and X-ray luminosities. It will allow for the first time statistically robust radio/X-ray correlation 
analyses of these quantities with relativistic jet parameters from VLBI observations. In addition, 
because of the ongoing MOJAVE monitoring observations, the Swift data will naturally yield to an 
unprecedented set of quasi-simultaneous broadband SED data from the radio, optical/UV and X-ray 
regime. 

\paragraph{New X-ray spectra for 36 MOJAVE sources:}
Our program started in November 2006. At the time of writing (March 2007), 94 objects have been observed. We concentrate here on
completed observations (i.e., $> \sim 10$\,ksec) of sources from the statistically complete MOJAVE
sample of the 133 radio-brightest, compact AGN in the northern sky. 
The XRT data were acquired mainly in Photon Counting (PC) mode.  We used the program {\sc XSELECT} to extract source 
and background counts from the cleaned event lists processed at Swift Science Data Center (SDC).  
The source spectrum is calculated from a circular region with the radius of 47\,arcsec, while the background region 
is selected as an annulus of the outer radius of 150\,arcsec and the inner radius of 70\,arcsec.  A pile-up correction 
is applied when the count rate exceeds 0.6 counts/s by excluding the central area of 7\,arcsec radius.  
RLCDs are found to have comparably simple X-ray spectra, typically well-approximated by an
absorbed power law.
Therefore, we performed 
a spectral fit to the time-averaged spectra with an absorbed power-law model. 
As of March 2007, we have completed
the observations of 36 out of 66 previously unobserved MOJAVE sources.   
In Table~\ref{tab:a}, we report
for the first time the basic X-ray spectral characteristics of these objects in the $(0.3-10)$\,keV band.
Errors quoted are at the 90\% confidence level.

All 36 objects have been detected with the XRT in 10\,ksec exposures. In some cases of either very low
X-ray flux or peculiar spectral or temporal source behavior, we have obtained follow-up observations
to improve the photon statistics or to trace source variability over a longer time range.
We find an average value of $\Gamma_{\rm ave} = 1.8$, with a standard deviation of $0.2$
similar to values of $\sim$ 1.6--1.7, which are typically found in the old 2\,cm-X-Sample 
(\cite{Kad05} and in prep.). The absorbing column densities determined have been compared to
the expected Galactic absorption values from the LAB survey \cite{Kal05}. Significant excess absorption
was found in 9 sources ($< 10^{21}$\,cm$^{-2}$ in all cases; compare Table~\ref{tab:a}).

\paragraph{New hard X-ray detections of five blazars with BAT:}
%The Burst Alert Telescope (BAT) is the hard X-ray instrument on-board \textit{Swift}. With it's all-sky coverage and high
%sensitivity down to $\sim 10^{-11}$\,erg\,s$^{-1}$\,cm$^{-2}$, BAT is the ideal instrument to characterize a large blazar
%sample like MOJAVE in the important 14--195\,keV band. This band coincides for many blazars with the transition region of the spectrum
%between the synchrotron and inverse-Compton components. This portion of the SED is of particular interest but is also 
%observationally most challenging because of the local minimum of the spectral energy distribution (SED) in this regime.
%Previously applied blind-search techniques did preferentially find those blazars who are both bright and distant, so that 
%their strong inverse-Compton component is redshifted into the BAT band. Consequently, 
The Burst Alert Telescope (BAT) is the hard X-ray instrument on-board \textit{Swift} 
and is a continuously operating all-sky hard X-ray monitor and survey instrument \cite{Mar05}. 
Prior to the beginning of our program,
17 blazars were 
detected by BAT (Tueller et al., in prep.). These sources were found via ``blind search'' of the whole sky, using
a sliding-cell detection method ({\sc batcelldetect}), requiring at least $\sim 5 \sigma$ significance of each individual 
source.
We have conducted a search of the BAT data base, based on the first nine months of observations, 
aiming for those MOJAVE sources for which our \textit{Swift} XRT 
fill-in observations did yield
an extrapolated $(14-195)$\,keV flux above or close to
$\sim 10^{-11}$\,erg\,s$^{-1}$\,cm$^{-2}$.
This criterion was met for 19 objects which we assign a ranking number, according to the value of the
extrapolated flux (rank\,1 for the highest to rank\,19 for the lowest extrapolated flux).  
We found significant excess flux at the positions of five 
blazars: NRAO\,140 (Rank\,1, $F_{\rm extrapol.}=4.73 \times 10^{-11}$\,erg\,s$^{-1}$\,cm$^{-2}$, $5.1 \sigma$), 
PKS B\,1510$-$089 (Rank\,2, $F_{\rm extrapol.}=3.68 \times 10^{-11}$\,erg\,s$^{-1}$\,cm$^{-2}$, $4.0 \sigma$),
PKS B\,2126$-$158 (Rank\,3, $F_{\rm extrapol.}=3.40 \times 10^{-11}$\,erg\,s$^{-1}$\,cm$^{-2}$, $3.4 \sigma$),  
3C\,345 (Rank\,6, $F_{\rm extrapol.}=2.61 \times 10^{-11}$\,erg\,s$^{-1}$\,cm$^{-2}$, $3.7 \sigma$),
and PKS B\,1222+216 (Rank\,19, $F_{\rm extrapol.}=9.36 \times 10^{-12}$\,erg\,s$^{-1}$\,cm$^{-2}$, $3.8 \sigma$).
A deeper analysis of the hard X-ray properties of these blazars is currently being performed.
Tentative signals ($\lesssim 3 \sigma$) are found at the position of the 5th-ranked source PKS\,B\,1127$-$145,
the 10th-ranked source PKS\,B\,0403$-$132, and the 16th-ranked source PKS\,B\,2008$-$159.

The rank distribution demonstrates that our method preferentially picks up those blazars
that are just below the BAT detection threshold of a few times $10^{-11}$\,erg\,s$^{-1}$\,cm$^{-2}$.
The fact that a significant signal is registered at the position of the relatively faint source PKS\,B\,1222$+$216
but not from comparably bright objects like the 4th-ranked source PKS\,B\,0723$-$08 suggests that X-ray
variability and spectral breaks above 10\,keV may play an important role. We expect to detect more MOJAVE
sources with BAT in the future, in particular with increased sensitivity in later stages of the BAT mission. 

\begin{table}
\begin{tabular}{clrccccc}
\hline
   \tablehead{1}{c}{b}{Source\\IAU B\,1950}
  & \tablehead{1}{c}{b}{Alt.\\Name} 
  & \tablehead{1}{r}{b}{Exposure\\ \[[ksec]}
  & \tablehead{1}{c}{b}{Photon\\Index}
  & \tablehead{1}{c}{b}{$N_{\rm H}$\\ \[[$10^{22}$\,cm$^{-2}$]}
  & \tablehead{1}{c}{b}{$N_{\rm H, Gal}$\\ \[[$10^{22}$\,cm$^{-2}$]} 
  & \tablehead{1}{c}{b}{$\chi^2_{\rm red}$/dof\\}
  & \tablehead{1}{c}{b}{$F_{(0.3-10)\,{\rm keV}}$\\ \[[erg\,s$^{-1}$\,cm$^{-2}$]}  
\\
\hline
0016$+$731 &            &8950   &$1.75_{-0.21}^{+0.23} $& $0.25_{-0.07}^{+0.08}$ & 0.19 &0.59/22       &       $3.30 \times 10^{-12}$  \\
0202$+$149 &            &8983   &$1.92_{-1.02}^{+1.43} $& $<0.65_{}^{}         $ & 0.05 &0.74/3        &       $3.04 \times 10^{-13}$  \\
0224$+$671 &4C\,+67.05  &12112  &$1.89_{-0.34}^{+0.39} $& $0.58_{-0.19}^{+0.25}$ & 0.40 &0.44/9        &       $1.23 \times 10^{-12}$  \\
0336$-$019 &CTA\,26     &10653  &$1.94_{-0.27}^{+0.31} $& $0.08_{-0.06}^{+0.07}$ & 0.06 &1.27/11       &       $1.17 \times 10^{-12}$  \\
0403$-$132 &            &9503   &$1.56_{-0.13}^{+0.14} $& $0.06_{-0.03}^{+0.04}$ & 0.04 &0.78/37       &       $5.02 \times 10^{-12}$  \\
0529$+$075 &            &19537  &$1.86_{-0.18}^{+0.19} $& $0.30_{-0.18}^{+0.19}$ & 0.17 &1.36/31       &       $1.92 \times 10^{-12}$  \\
0529$+$483 &            &9124   &$1.28_{-0.63}^{+0.79} $& $<0.37_{}^{}         $ & 0.25 &0.87/2        &       $9.21 \times 10^{-13}$  \\
1124$-$186 &            &9515   &$1.99_{-0.36}^{+0.45} $& $0.07_{-0.06}^{+0.08}$ & 0.04 &1.26/8        &       $9.71 \times 10^{-13}$  \\
1150$+$812 &            &11244  &$1.74_{-0.39}^{+0.45} $& $<0.07_{}^{}         $ & 0.05 &1.24/6        &       $8.48 \times 10^{-13}$  \\
1324$+$224 &            &13330  &$1.78_{-0.17}^{+0.19} $& $0.07_{-0.04}^{+0.05}$ & 0.02 &0.86/24       &       $2.01 \times 10^{-12}$  \\
1417$+$385 &            &11730  &$2.11_{-0.41}^{+0.52} $& $0.08_{-0.07}^{+0.10}$ & 0.01 &0.46/5        &       $5.56 \times 10^{-13}$  \\
1504$-$167 &            &12571  &$1.80_{-0.54}^{+0.70} $& $<0.22_{}^{}         $ & 0.07 &0.64/7        &       $3.34 \times 10^{-13}$  \\
1538$+$149 &4C\,+14.60  &9527   &$2.28_{-0.40}^{+0.49} $& $0.12_{-0.07}^{+0.09}$ & 0.03 &0.60/11       &       $1.17 \times 10^{-12}$  \\
1546$+$027 &            &9657   &$1.76_{-0.16}^{+0.17} $& $0.07_{-0.04}^{+0.04}$ & 0.07 &1.03/31       &       $3.45 \times 10^{-12}$  \\
1606$+$106 &4C\,+10.45  &49342  &$1.43_{-0.09}^{+0.10} $& $0.04_{-0.02}^{+0.02}$ & 0.04 &0.93/73       &       $1.92 \times 10^{-12}$  \\
1611$+$343 &DA\,406     &9789   &$1.77_{-0.27}^{+0.34} $& $<0.11_{}^{}         $ & 0.01 &0.74/9        &       $1.18 \times 10^{-12}$  \\
1637$+$574 &            &10140  &$1.91_{-0.13}^{+0.14} $& $0.05_{-0.03}^{+0.03}$ & 0.01 &0.90/40       &       $3.95 \times 10^{-12}$  \\
1638$+$398 &NRAO\,512   &12267  &$1.49_{-0.23}^{+0.29} $& $<0.04_{}^{}         $ & 0.01 &1.17/5        &       $7.58 \times 10^{-13}$  \\
1726$+$455 &            &10722  &$2.04_{-0.31}^{+0.36} $& $0.15_{-0.07}^{+0.08}$ & 0.02 &1.23/12       &       $1.20 \times 10^{-12}$  \\
1730$-$130 &NRAO\,530   &9912   &$1.46_{-0.29}^{+0.32} $& $0.29_{-0.11}^{+0.14}$ & 0.18 &0.98/12       &       $2.18 \times 10^{-12}$  \\
1739$+$522 &4C\,+51.37  &10304  &$1.55_{-0.20}^{+0.23} $& $0.07_{-0.05}^{+0.06}$ & 0.03 &0.72/18       &       $2.30 \times 10^{-12}$  \\
1741$-$038 &            &9927   &$1.72_{-0.31}^{+0.34} $& $0.41_{-0.14}^{+0.18}$ & 0.18 &0.72/12       &       $2.01 \times 10^{-12}$  \\
1751$+$288 &            &9772   &$1.77_{-0.39}^{+0.45} $& $0.12_{-0.09}^{+0.11}$ & 0.05 &0.83/7        &       $1.10 \times 10^{-12}$  \\
1758$+$388 &            &9711   &$2.14_{-0.90}^{+1.30} $& $<0.38_{}^{}         $ & 0.03 &0.70/4        &       $2.84 \times 10^{-13}$  \\
1800$+$440 &            &10098  &$1.75_{-0.16}^{+0.17} $& $0.04_{-0.03}^{+0.03}$ & 0.03 &0.98/31       &       $3.36 \times 10^{-12}$  \\
1849$+$670 &            &9952   &$2.07_{-0.28}^{+0.31} $& $0.17_{-0.07}^{+0.08}$ & 0.05 &1.65/16       &       $1.71 \times 10^{-12}$  \\
1936$-$155 &            &9929   &$2.25_{-0.49}^{+0.60} $& $0.24_{-0.12}^{+0.17}$ & 0.08 &0.50/5        &       $7.15 \times 10^{-13}$  \\
1958$-$179 &            &8997   &$1.84_{-0.27}^{+0.29} $& $0.16_{-0.07}^{+0.09}$ & 0.07 &0.90/14       &       $1.95 \times 10^{-12}$  \\
2005$+$403 &            &11986  &$1.69_{-0.27}^{+0.29} $& $0.50_{-0.15}^{+0.19}$ & 0.48 &0.88/15       &       $1.98 \times 10^{-12}$  \\
2008$-$159 &            &9657   &$1.75_{-0.12}^{+0.13} $& $0.14_{-0.03}^{+0.04}$ & 0.06 &1.08/48       &       $5.71 \times 10^{-12}$  \\
2021$+$317 &4C\,+31.56  &28861  &$1.96_{-0.65}^{+1.04} $& $1.21_{-0.76}^{+0.95}$ & 0.52 &1.98/4        &       $3.24 \times 10^{-13}$  \\
2021$+$614 &            &17562  &$0.87_{-0.88}^{+1.26} $& $<2.01_{}^{}         $ & 0.14 &0.51/3        &       $2.14 \times 10^{-13}$  \\
2037$+$511 &3C\,418     &10189  &$1.70_{-0.38}^{+0.45} $& $0.66_{-0.25}^{+0.37}$ & 0.54 &0.91/11       &       $2.02 \times 10^{-12}$  \\
2136$+$141 &OX\,161     &10140  &$1.43_{-0.25}^{+0.28} $& $<0.15_{}^{}         $ & 0.06 &0.48/11       &       $1.67 \times 10^{-12}$  \\
2201$+$171 &            &9585   &$1.85_{-0.30}^{+0.34} $& $0.10_{-0.07}^{+0.09}$ & 0.05 &0.89/7        &       $9.80 \times 10^{-13}$  \\
2216$-$038 &            &9650   &$1.82_{-0.29}^{+0.33} $& $0.07_{-0.06}^{+0.08}$ & 0.06 &1.16/12       &       $1.53 \times 10^{-12}$  \\      
\hline
\end{tabular}
\caption{X-ray spectral fitting parameter of MOJAVE sources}
\label{tab:a}
\end{table}                    

%%%%%%%%%%%%%%%%%%%%%%%%%%%%%%%%%%%%%%%%%%%%%%%%
%% BACKMATTER
%%%%%%%%%%%%%%%%%%%%%%%%%%%%%%%%%%%%%%%%%%%%%%%%

\begin{theacknowledgments}
The authors wish to acknowledge the efforts and contributions of the \textit{Swift} BAT and MOJAVE teams.
\end{theacknowledgments}


%\bibliography{bibtex}

\end{document}